\documentclass[11pt, letterpaper]{amsart}
\usepackage{graphicx, amssymb}

\newtheorem{thm}{Theorem}[section]
\newtheorem{cor}[thm]{Corollary}

\newtheorem{prop}[thm]{Proposition}
\theoremstyle{definition}

\newtheorem{ass}[thm]{Assumption}
\newtheorem{asss}[thm]{Assumptions}
\theoremstyle{remark}
\newtheorem{rem}[thm]{Remark}

\numberwithin{equation}{section}

\newcommand{\set}[1]{\left\{#1\right\}}
\newcommand{\Real}{\mathbb R}
\newcommand{\Natural}{\mathbb N}

\newcommand{\B}{\mathcal{B}}
\newcommand{\such}{\, | \,}
\newcommand{\bigsuch}{\ \Big| \ }
\newcommand{\limn}{\lim_{n \to \infty}}

\newcommand{\dfn}{\, := \,}

\newcommand{\prob}{\mathbb{P}}

\newcommand{\Leb}{\mathsf{Leb}}

\newcommand{\expec}{\mathbb{E}}

\newcommand{\basis}{(\Omega, \mathcal{F}, \mathbf{F}, \prob)}

\newcommand{\F}{\mathcal{F}}

\newcommand{\cadlag}{c\`adl\`ag}

\newcommand{\ud}{\, \mathrm d}

\newcommand{\inner}[2]{\left \langle #1 , \, #2 \right \rangle}
\newcommand{\innerw}[2]{\left \langle #1 , #2 \right \rangle}
\newcommand{\num}{num\'eraire}
\newcommand{\X}{\mathcal{X}}

\newcommand{\hX}{\widehat{X}}
\newcommand{\hL}{\widehat{L}}

\newcommand{\g}{\mathfrak{g}}

\newcommand{\pare}[1]{\left(#1\right)}
\newcommand{\bra}[1]{\left[#1\right]}
\newcommand{\dbra}[1]{[\kern-0.15em[ #1 ]\kern-0.15em]}
\newcommand{\dbraco}[1]{[\kern-0.15em[ #1 [\kern-0.15em[}
\newcommand{\C}{\mathfrak{C}}

\newcommand{\bF}{\mathbf{F}}

\newcommand{\indic}{\mathbb{I}}

\newcommand{\absco}{{<\kern-0.53em<}}

\newcommand{\xhat}{{\widehat{X}}}

\newcommand{\Op}{\mathcal{O}}
\newcommand{\TOp}{\mathcal{T}}

\newcommand{\ito}{It\^o}

\newcommand{\I}{\mathfrak{I}}
\newcommand{\Cc}{\check{\mathfrak{C}}}

\newcommand{\Lhat}{\widehat{L}}

\newcommand{\cd}{c^{\dagger}}
\begin{document}

\pagestyle{plain}

\title[Minimizing the expected market time to reach a certain wealth level]{Minimizing the expected market time to reach a certain wealth level}%
\author{Constantinos Kardaras}%
\address{Constantinos Kardaras, Mathematics and Statistics Department, Boston University, 111 Cummington Street, Boston, MA 02215, USA.}%
\email{kardaras@bu.edu}%

\author{Eckhard Platen}%
\address{Eckhard Platen, School of Finance and Economics \& Department of Mathematical Sciences,
University of Technology, Sydney, P.O. Box 123, Broadway, NSW 2007, Australia.}%
\email{eckhard.platen@uts.edu.au}%

\subjclass[2000]{60H99, 60G44, 91B28, 91B70}
\keywords{Num\'eraire portfolio, growth-optimal portfolio, market time, upcrossing, overshoot, exponential L\'evy markets, It\^o markets, semimartingale markets}%

\date{\today}%
\begin{abstract}
In a financial market model, we consider variations of the problem of minimizing the expected time to upcross a certain wealth level. For exponential L\'evy markets, we show the asymptotic optimality of the growth-optimal portfolio for the above problem and obtain tight bounds for the value function for any wealth level. In an \ito \ market, we employ the concept of \textsl{market time}, which is a clock that runs according to the underlying market growth. We show the optimality of the growth-optimal portfolio for minimizing the expected market time to reach any wealth level. This reveals a general definition of market time which can be useful from an investor's point of view. We utilize this last definition to extend the previous results in a general semimartingale setting.
\end{abstract}

\maketitle

\setcounter{section}{0}

\section{Introduction}

The problem of \emph{quickly} reaching certain goals in wealth management is one of the most fundamental tasks in the theory and practice of finance. However, making this idea mathematically precise  has been a challenge. In particular, this would require a quantification of what is meant by achieving goals ``quickly'' in a model-independent manner, or, even better, coming endogenously from the description of the market as is perceived by its participants. Such a mathematically precise description of the flow of time, as well as the corresponding optimal investment strategy, is clearly valuable. If a robust, model-independent answer to the previous questions can be given, it would go a long way towards a better understanding of the problem, as its statement should provide a deep inside into key quantitative characteristics of the market. Our aim in this paper is to present a way of addressing the aforementioned issues.

\smallskip

We proceed with a more thorough description of the problem. Imagine an investor holding some minute capital-in-hand, aiming to reach as quickly as possible a substantial wealth level by optimally choosing an investment opportunity in an active market. No matter what the mathematical formalization of the objective is, as long as it reasonably describes the above informal setting, intuition suggests that the investor should pick an aggressive strategy that provides ample wealth growth. The most famous wealth-optimizing strategy that could potentially achieve this is the \emph{growth-optimal} strategy, which is sometimes also called \emph{Kelly} strategy, as the latter was introduced in \cite{MR0090494}. Therefore, the portfolio generated by the growth-optimal strategy is a strong candidate for solving the aforementioned problem, at least in an approximate sense. This last point is augmented by the long line of research on the importance and optimality properties of the growth optimal portfolio; we mention for example the very incomplete list: \cite{Long90}, \cite{MR929084}, \cite{MR1849424}, \cite{MR2194899}, \cite{MR2284490}, \cite{MR2335830}. Note also that minimizing expected time to reach a wealth level is not the only interesting objective that one can seek. For example, maximizing the probability that a wealth level will be reached before some future time is also interesting; in this respect, see  \cite{MR1724568}, \cite{MR1842286}.

Here, we shall identify a variant of the ``quickest goal reach'' problem for \emph{continuous-time} models where the growth-optimal portfolio is indeed the best. The problem we consider then is that of minimizing the expected \emph{market time} that it will take to reach a certain wealth level. Market time will be defined as a natural time scale which runs fast when the compensation for taking risk in the market is high and vice-versa. In a market with continuous asset prices, this will be achieved  by setting the slope of the market time equal to half the squared \emph{risk premium}. In this case, it equals the growth rate of the corresponding growth-optimal portfolio, which leads to the interpretation of market time as integrated maximum growth rate.

\smallskip

The first attempt to minimize the expected upcrossing time in a discrete-time gambling-system model was described in \cite{MR0135630}, where indeed the \emph{near} optimal wealth process was found to be characterized by Kelly's growth-optimal strategy. Models of gambling systems, as considered in \cite{MR0135630}, could be interpreted as discrete-time financial markets where the log-asset-price processes are random walks with a finite number of possible values for the increment of each step. The natural continuous-time generalization of the above setting is to consider \textsl{exponential L\'evy markets}, i.e., markets where the log-asset-price processes have independent and stationary increments. For these markets, we establish here the exact analogues of the results in \cite{MR0135630}.

A continuous-time problem in the context of a Black-Scholes market was treated in \cite{HeSu}, and then as an application of a more abstract problem in \cite{MR872458}, using essentially methods of dynamic programming. In this case, \textsl{the \num \ portfolio} of the market, which was introduced in \cite{Long90} and is also called \textsl{the growth-optimal portfolio} as it is generated by the analogue of Kelly's growth-optimal strategy, is truly optimal for minimizing the expected calendar time to reach any wealth level. Unfortunately, the moment that one considers more complex It\^o-process models, for example ones that are modelling feedback effects, as the leverage effect in \cite{Black76}, the growth-optimal portfolio is no longer optimal for the problem of minimizing expected calendar time for upcrossing a certain wealth level. In fact, for general non-Markovian models there does not seem to be any hope in identifying what the optimal strategy and wealth process are when minimizing expected calendar time. We note however that for \emph{Markovian} models one can still characterize the optimal strategy and portfolio in terms of a Hamilton-Jacobi-Bellman equation, which will most likely then have to be solved numerically.

\smallskip

We introduce in this paper a \emph{market clock} which does not count time according to the natural calendar flow, but rather according to the overall market growth. Under the objective that one minimizes expected market time, we show here that the solution again yields the growth-optimal portfolio as nearly optimal. There is a slight problem that results in the non-optimality of the growth-optimal portfolio, if for finite wealth levels some \emph{overshoot} is possible over the targeted wealth level at the time of the upcrossing. If there is no overshoot, which happens in particular in models with continuous asset prices, then the growth-optimal portfolio is indeed optimal. In \cite{MR520737}, the author considers a ramification of the problem by offering a \emph{rebate} for the overshoot that results in the growth-optimal portfolio being again optimal. Of course, we could do this even in the most general case. Since this rebate inclusion is somewhat arbitrary, we shall refrain from using it in our own analysis.

The optimality of the growth-optimal portfolio for minimizing expected time according to a clock counting time according to the overall market growth sounds a bit like a tautological statement. However, we shall make a conscious effort to convey that the concept of market time is very natural, by taking a stepwise approach in the model generality that we consider. The exponential L\'evy process case is considered first. There, the market-time flow coincides with the calendar-time flow up to a multiplicative constant, since the model coefficients remain constant through time. As soon as the model coefficients are allowed to randomly change, one can regard the passage of time in terms of the \emph{opportunities} for profit that are available. We first discuss this in the realm of markets where asset-prices are modeled via It\^o processes, where the arguments are more intuitive. As soon as the natural candidate for the market time is understood, we proceed to discuss the results in the very general semimartingale model.

\smallskip

The results presented in this work are generalizations of the constant-coefficient result in \cite{HeSu}. The use of martingale methods and a natural definition of market time that we utilize make the proof of our claims more transparent and widens the scope and validity of the corresponding statements.

%
%

\bigskip

The structure of the paper is as follows. In Section \ref{sec: descr of prob} we introduce the general financial market model, we define the problem of minimizing expected market time and present the standing assumptions, which are basically the existence of the \num \ portfolio. In Section \ref{sec: ELM} we specialize in the case of exponential L\'evy market models, where market time and calendar time coincide up to a multiplicative constant. Our first main result gives tight bounds for the near-optimal performance of the growth-optimal portfolio for any wealth level, that also result in its asymptotic optimality for increasing wealth levels. In Section \ref{sec: ito} we use It\^o processes to  model the market. After some discussion on the concept of market time, our second main result shows also here the optimality of the growth-optimal portfolio. In Section \ref{sec: ZMPJR}, the concept of market time in a general semimartingale setting is introduced and a general result that covers all previous cases is presented. Finally, Section \ref{sec: proof} contains the proofs of the results in the previous sections.

\section{Description of the Problem} \label{sec: descr of prob}

In the following general remarks we fix some notation that will be used throughout.

By $\Real_+$ we shall denote the positive real line, $\Real^d$ the $d$-dimensional Euclidean space, and $\Natural$ the set of natural numbers $\{ 1, 2, \ldots \}$. Superscripts will be used to indicate coordinates, both for vectors and for processes; for example $z \in \Real^d$ is written $z = (z^1, \ldots, z^d)$. On $\Real^d$, $\inner{\cdot}{\cdot}$ will denote the \emph{usual} inner product: $\inner{y}{z} \dfn \sum_{i=1}^d y^i z^i$ for $y$ and $z$ in $\Real^d$. Also $|\cdot|$ will denote the usual norm: $|z| \dfn \sqrt{\inner{z}{z}}$ for $z \in \Real^d$.

On $\Real_+$ equipped with the Borel $\sigma$-field $\B(\Real_+)$, $\Leb$ will denote the \textsl{Lebesgue measure}.

All stochastic processes appearing in the sequel are defined on a filtered probability space $(\Omega, \, \F, \, \bF, \, \prob)$. Here, $\prob$ is a probability on $(\Omega, \F)$, where $\F$ is a $\sigma$-algebra that will make all involved random variables measurable. The filtration $\bF = (\F_t)_{t \in \Real_+}$ is assumed to satisfy the \textsl{usual hypotheses} of right-continuity and saturation by $\prob$-null sets. It will be assumed throughout that $\F_0$ is trivial modulo $\prob$.

For a \cadlag \ (right continuous with left limits) stochastic process $X = (X_t)_{t \in \Real_+}$, define $X_{t-} \dfn \lim_{s \uparrow t} X_s$ for $t > 0$ and $X_{0 -} \dfn 0$. The process $X_-$ will denote this last left-continuous version of $X$ and $\Delta X \dfn X- X_-$ will be the jump process of $X$.

\subsection{Assets and wealth processes}

The $d$-dimensional semimartingale $S = (S^1, \ldots, S^d)$ will be denoting the \emph{discounted}, with respect to the savings account, price process of $d$ financial assets.

Starting with initial capital $x \in \Real_+$, and investing according to some predictable and $S$-integrable strategy $\vartheta$, an investor's \emph{discounted} total wealth process is given by
\begin{equation} \label{eq: wealth process}
X^{x, \vartheta} \dfn x + \int_0^\cdot \innerw{\vartheta_t}{\ud S_t}.
\end{equation}

Reflecting the investor's ability only to hold a portfolio of nonnegative total tradeable wealth, we then define the set of all nonnegative wealth processes starting from initial capital $x \in \Real_+$:
\[
\X (x) \dfn \set{ X^{x, \vartheta} \text{ as in } \eqref{eq: wealth process} \bigsuch \vartheta \text{ is predictable and } S\text{-integrable, and } X^{x, \vartheta} \geq 0 }.
\]
It is straightforward that $\X(x) = x \X(1)$ and that $x \in \X(x)$ for all $x \in \Real_+$. We also set $\X := \bigcup_{x \in \Real_+} \X(x)$.

\subsection{The problem} \label{subsec: the problem}

We shall be concerned with the problem of \emph{quickly reaching a wealth level $\ell$ starting from capital $x$}. This, of course, is nontrivial only when $x < \ell$, which will be tacitly assumed throughout. The challenge is now to rigorously define what is meant by ``quickly''.
Take $\Op = (\Op_t)_{t \in \Real_+}$ to be an increasing and adapted process such that, $\prob$-a.s., $\Op_0 = 0$ and $\Op_\infty = + \infty$. $\Op$ will be representing some kind of internal clock of the market, which we shall call \textsl{market time}. In the following sections we shall  be more precise on choosing $\Op$, guided by what we shall learn when identifying the consequences of applying the growth-optimal strategy.

For any \cadlag \ process $X$ and $\ell \in \Real_+$, define the \textsl{first upcrossing market time of $X$ at level $\ell$}:
\begin{equation} \label{eq: mark upcross defn}
\TOp (X; \ell) \dfn \inf \set{\Op_t \in \Real_+ \such X_t \geq \ell}.
\end{equation}
Of course, if $\ell \leq x$ then $\TOp(X; \ell) = 0$ for all $X \in \X(x)$.
With the aforementioned inputs, define for all $x < \ell$ the value function
\begin{equation} \label{eq: value function}
v(x; \ell) \dfn \inf_{X \in \X(x)} \expec \bra{\TOp (X; \ell) }.
\end{equation}

Our aims in this work are to:
\begin{itemize}
  \item identify a natural definition for the market time $\Op$;
  \item obtain an explicit formula, or at least some useful tight bounds, for the value function $v(x; \ell)$ of \eqref{eq: value  function}; and
  \item find the optimal, or perhaps \emph{near} optimal, portfolio for the above problem.
\end{itemize}

\subsection{Standing assumptions}
In order to make headway with the problem described in \S \ref{subsec: the problem}, we shall make two natural and indispensable assumptions regarding the financial market that will be in force throughout.

\begin{asss} \label{ass: general}
In our financial market model, we assume the following:

\begin{enumerate}
  \item There exists $\hX \in \X(1)$ such that $X / \hX$ is a supermartingale for all $X \in \X$.
  \item For every $\ell \in \Real_+$, there exists $X \in \X(1)$, possibly depending on $\ell$, such that, $\prob$-a.s., $\TOp(X; \ell) < + \infty$.
\end{enumerate}
\end{asss}

A process $\hX$ with the properties described in Assumption \ref{ass: general}(1) is unique and is called the \textsl{\num \ portfolio}. Existence of the \num \ portfolio is a \emph{minimal} assumption for the viability of the financial market. It is essentially equivalent to the boundedness in probability of the set $\set{X_T \such X \in \X(1)}$ of all possible discounted wealths starting from unit capital and observed at any time $T \in \Real_+$. We refer the interested reader to \cite{MR2284490}, \cite{MR2335830} and \cite{KarPla07} for more information in this direction.
We shall frequently refer to the \num \ portfolio as the \emph{growth-optimal portfolio}, as the two notions coincide.

Assumption \ref{ass: general}(2) constitutes what has been  coined a ``favorable game'' in \cite{MR0135630} and it is \emph{necessary} in order for the problem described in \eqref{eq: value function} to have finite value  and therefore to be well-posed. Under Assumption \ref{ass: general}(2), and in view of the property $\X(x) = x \X(1)$ for $x \in \Real_+$, it is obvious that for all $x \in \Real_+$ and $\ell \in \Real_+$, there exists $X \in \X(x)$ such that $\prob \bra{\TOp(X; \ell) < + \infty} = 1$.

Actually, if Assumption \ref{ass: general}(1) is in force, Assumption \ref{ass: general}(2) has a convenient equivalent.

\begin{prop} \label{prop: if 1 then num is the one}
Under Assumption \ref{ass: general}(1), Assumption \ref{ass: general}(2) is equivalent to:
\begin{enumerate}
  \item[$(2')$] $\lim_{t \to + \infty} \hX_t = + \infty$, $\prob$-a.s.
\end{enumerate}
\end{prop}

This last result enables one to check easily the validity of Assumptions \ref{ass: general} by looking only at the \num \ portfolio. In each of the specific cases we shall consider in the sequel, equivalent characterizations of Assumptions \ref{ass: general} will be given in terms of the model under consideration.

\section{Exponential L\'evy Markets} \label{sec: ELM}

\subsection{The set-up}

For this section we assume that the discounted asset-price processes satisfy $\ud S^i_t = S^i_{t-} \ud R^i_t$ for $t \in \Real_+$, where, for all $i = 1, \ldots, d$, $R^i$ is a L\'evy process on $\basis$. Each $R^i$ for $i=1, \ldots, d$ is the \textsl{total returns process} associated to $S^i$.

In order to make sure that the asset-price processes remain nonnegative, it is necessary and sufficient that $\Delta R^i \geq -1$ for all $i = 1, \ldots, d$. We shall actually impose a further restriction on the structure of the jumps of the returns processes, also bounding them from above. This is mostly done in order to obtain later in Theorem \ref{thm: main of ELM} a statement which parallels the result in \cite{MR0135630}. For the asymptotic result that will be presented in \S \ref{subsec: proof of main ito} this bounded-jump assumption will be dropped.

\begin{ass} \label{ass: bdd levy meas}
For all $i = 1, \ldots, d$ we have $-1 \leq \Delta R^i \leq \kappa$, for some $\kappa \in \Real_+$.
\end{ass}

Denote by $R$ the $d$-dimensional L\'evy process $(R^1, \ldots, R^d)$. In view of the boundedness of the jumps of $R$, as stated in Assumption \ref{ass: bdd levy meas} above, we can write
\begin{equation} \label{eq: canonical representation}
R_T = a T + \sigma W_T + \int_{[0, T] \times \Real^d}  z \pare{ \mu(\ud z, \ud t) - \nu (\ud z) \ud
t }
\end{equation}
for all $T \in \Real_+$. In view of Assumption \ref{ass: bdd levy meas}, the elements in the above representation satisfy:
\begin{itemize}
  \item $a \in \Real^d$.
  \item $\sigma$ is a $(d \times m)$-matrix, where $m \in \Natural$.
  \item $W$ is a standard $m$-dimensional Brownian motion on $\basis$.
  \item $\mu$ is the \textsl{jump measure} of $R$, i.e., the random
counting measure on $\Real_+ \times \Real^d$ defined via $\mu ([0,T] \times E) := \sum_{0 \leq t
\leq T} \indic_{E \setminus \{0\}}(\Delta R_t)$ for $T \in \Real_+$ and $E \subseteq
\Real^d$.
  \item $\nu$, the \emph{compensator} of $\mu$, is a \textsl{L\'evy measure} on $(\Real^d, \B(\Real^d))$, where $\B(\Real^d)$ is the Borel $\sigma$-field on $\Real^d$. More precisely, $\nu$ is a measure with $\nu[\set{0}] = 0$, $\nu \bra{\Real^d \setminus [-1, \kappa]} = 0$ and $\int_{\Real^d} |x|^2 \nu [\ud x] < + \infty$.
\end{itemize}
For more information on L\'evy processes one can check for example \cite{MR1739520}.

Define the $(d \times d)$ matrix $c \dfn \sigma \sigma^\top$, where ``$\top$'' denotes matrix transposition. The triplet $(a, \, c, \, \nu)$ will play a crucial role in the discussion below.

\smallskip

In the notation of \eqref{eq: wealth process}, let $X^{x, \vartheta} \in \X(x)$. The nonnegativity requirement $X^{x, \vartheta} \geq 0$ is equivalent to $\Delta X^{x, \vartheta} \geq X_-^{x, \vartheta}$, or further to $\inner{\vartheta }{\Delta S} \geq X_-^{x, \vartheta}$. Since $\Delta S^i = S^i_- \Delta R^i$ for each $i = 1, \ldots, d$, and recalling that $\nu$ is the L\'evy measure of $R$, we conclude that $X^{x, \vartheta} \geq 0$ if and only if
\[
\pare{\vartheta_t^i(\omega) S^i_{t-} (\omega)}_{i = 1, \ldots, d} \in X^{x ,\vartheta}_{t-} (\omega) \C, \text{ for all } (\omega, t) \in \Omega \times \Real_+,
\]
where $\C$ is the set of \textsl{natural constraints} defined via
\[
\C \dfn \set{ \eta \in \Real^d \bigsuch  \nu \big[ z \in \Real^d \such \innerw{\eta}{z
} < -1 \big] = 0  }.
\]
It is easy to see that $\C$ is convex; it is also closed, as follows from Fatou's lemma.

\subsection{Growth rate}
For any $\pi \in \C$, define
\begin{equation} \label{eq: growth rate}
\g(\pi) \dfn \innerw{\pi}{a} - \frac{1}{2} \innerw{\pi}{c \pi} - \int_{\Real^d} \bra{\innerw{\pi}{z} - \log(1 + \innerw{\pi}{z}) } \nu [\ud z].
\end{equation}
For $\pi \in \C$, $\g(\pi)$ is the drift rate of the logarithm of the wealth process $X \in \X(1)$ that satisfies $\ud X_t = X_{t-} \innerw{\pi}{\ud R_t} = X_{t-} \ud \innerw{\pi}{R_t}$ for all $t \in \Real_+$; for this reason, $\g(\pi)$ is also called the \textsl{growth rate} of the last wealth process.

Define $\g^* \dfn \sup_{\pi \in \C} \g(\pi)$ to be the maximum growth rate. Since $0 \in \C$, we certainly have $\g^* \geq \g(0) = 0$. Actually, under the bounded-jump Assumption \ref{ass: bdd levy meas}, the standing Assumptions \ref{ass: general} are equivalent to $0 < \g^* < \infty$. In order to achieve this last claim, we shall connect the viability of the market with the concept of immediate arbitrage opportunities, as will be now introduced.

\subsection{Market viability}

Define the set $\I$ of \textsl{immediate arbitrage opportunities} to consist of all vectors $\xi \in \Real^d$ such that $c \xi = 0$, $\nu \big[ z \in \Real^d \such \innerw{\xi}{z
} < 0 \big] = 0$  and  $\inner{\xi}{a} \geq 0$, and where further at least one of $\nu \big[ z \in \Real^d \such \innerw{\xi}{z
} > 0 \big] > 0$  or $\inner{\xi}{a} > 0$ holds. As part of the next result, we get that the previously-described exponential L\'evy market is viable if and only if the intersection of $\I$ with the \textsl{recession cone} of $\C$, defined as $\check{\C} \dfn \bigcap_{u > 0} u \C$, is empty.

\begin{prop} \label{prop: nec and suf ELM}
Assumptions \ref{ass: general} are equivalent to requiring both $\I \cap \Cc = \emptyset$ and $\g^* > 0$.

Suppose now that the above is true, as well as that Assumption \ref{ass: bdd levy meas} is in force. Then, $\g^* < \infty$ and there exists $\rho \in \C$ such that $\g(\rho) = \g^*$. Furthermore, the \num \ portfolio $\hX$ satisfies the dynamics $\ud \hX_t = \hX_{t-} \innerw{\rho}{\ud R_t} = \hX_{t-} \ud \innerw{\rho}{R_t}$. In other words, for $T \in \Real_+$,
\begin{equation} \label{eq: num ELM}
\log \big( \hX_T \big) = \innerw{\rho}{R_T} - \frac{1}{2} \innerw{\rho}{c \rho} T - \sum_{0 \leq t \leq T} \pare{\innerw{\rho}{\Delta R_t} - \log \pare{1 + \innerw{\rho}{\Delta R_t}}}.
\end{equation}
\end{prop}

Instead of using the general Assumptions \ref{ass: general} in this section, we shall use the equivalent conditions $\I \cap \Cc = \emptyset$ and $\g^* > 0$. We also note that the vector $\rho \in \C$ in the statement of Proposition \ref{prop: nec and suf ELM} that leads to the \num \ portfolio is \emph{essentially} unique, modulo any degeneracies that might be present in the market and lead to non-zero portfolios having zero returns.

\subsection{The main result}

Since L\'evy processes have stationary and independent increments, the natural candidate for market time is to consider calendar time up to a multiplicative constant $\gamma > 0$, i.e., to set $\Op_t = \gamma t$ for $t \in \Real_+$. In Theorem \ref{thm: main of ELM} below, we shall actually choose $\gamma = \g^*$. This turns out to be the appropriate choice of market velocity that reflects a universal characteristic of the market and will result in the bounds \eqref{eq: bounds ELM} for the optimal upcrossing time in Theorem \ref{thm: main of ELM} below  not to depend on the actual model under consideration.

\begin{thm} \label{thm: main of ELM}
We work under Assumption \ref{ass: bdd levy meas}, and also assume that $\I \cap \Cc = \emptyset$ and $\g^* > 0$. Define the finite nonnegative constant $\alpha \dfn \inf \set{\beta \in \Real_+ \such \nu \bra{z \in \Real^d \such \inner{\rho}{z} > \beta} = 0}$. Let the market time $\Op$ be defined via $\Op_t = \g^* t$ for all $t \in \Real_+$. With $\xhat(x) \dfn x \xhat$, we have the inequalities:
\begin{equation} \label{eq: bounds ELM}
\log \pare{\frac{\ell}{x}} \ \leq \ v(x; \ell) \ \leq \ \expec \big[ \TOp(\xhat(x); \ell) \big] \ \leq \ \log \pare{\frac{\ell}{x}} + \log(1 + \alpha).
\end{equation}
\end{thm}

Actually, Theorem \ref{thm: main of ELM} is an instance of a more general statement that will be presented in Section \ref{sec: ZMPJR}. We note that the bounds \eqref{eq: bounds ELM} are in complete accordance with the discrete-time result in \cite{MR0135630} and that the nonnegative constant $\log(1 + \alpha)$ does not involve $x$ or $\ell$.

\begin{rem}
Under a mild condition, namely that the marginal one-dimensional distributions of $\log (\hX)$ are non-lattice, the overshoot of $\log (\hX)$ over the level $\log(\ell)$ actually has a limiting distribution as $\ell \to \infty$ that is supported on $[0, \log(1 + \alpha)]$. In that case,
\[
\lim_{\ell \to \infty} \pare{\expec \big[ \TOp(\xhat(x); \ell) \big] - \log \pare{\frac{\ell}{x}}}
\]
exists and is exactly equal to the mean of that limiting distribution.
\end{rem}

\subsection{True optimality}
There is a special case when the growth-optimal portfolio is indeed optimal for all levels $\ell$, which covers in particular the Black-Scholes market result in \cite{HeSu}. The following result directly stems out of the statement of Theorem \ref{thm: main of ELM}.

\begin{cor} \label{cor: no neg jumps in num of ELM}
Suppose that the \num \ portfolio $\hX$ of \eqref{eq: num ELM} has no positive jumps: $\inner{\rho}{\Delta R} \leq 0$. Then,
\[
v(x; \ell) \ = \ \log \pare{\frac{\ell}{x}} \ = \ \expec \big[ \TOp(\xhat(x); \ell) \big].
\]
\end{cor}

For an easy example where the last equality occurs, consider in \eqref{eq: canonical  representation} the case where $d=1$, $\kappa = 0$ and $a = a^1 > 0$. This is a reasonable model where the excess rate of return is strictly positive and only negative jumps are present in the dynamics of the discounted asset-price process.

\subsection{Asymptotic optimality without the bounded-jump assumption}
\label{subsec: asympt opt ELM}
Theorem \ref{thm: main of ELM} gives the asymptotic (for large $\ell$) optimality of the growth-optimal portfolio, since, by \eqref{eq: bounds  ELM},
\begin{equation} \label{eq: asymptotic ELM}
\lim_{\ell \to \infty} \frac{v(x; \ell)}{\log (\ell)} \ = \ 1 \ = \ \lim_{\ell \to \infty} \frac{\expec \big[ \TOp(\xhat(x); \ell) \big]}{\log (\ell)}.
\end{equation}

The validity of the asymptotic optimality in \eqref{eq:  asymptotic ELM} goes well-beyond the bounded-jump Assumption \ref{ass: bdd levy meas}, as we shall describe now. For the total returns process $R = (R^1, \ldots, R^d)$, we can write the canonical representation \eqref{eq: canonical representation} if and only if the L\'evy measure $\nu$ is such that $\int_{\Real^d} \pare{|x| \wedge |x|^2} \nu [\ud x] < + \infty$. In that case, the definition in \eqref{eq: growth  rate} of the growth rate is still the same, even without the validity of Assumption \ref{ass: bdd levy  meas}. We then have the following result.

\begin{prop} \label{prop: asymptotic ELM}
Suppose that the canonical representation \eqref{eq: canonical representation} is valid. Then, if $\I \cap \Cc = \emptyset$ and $\g^* > 0$ hold, we have $\g^* < \infty$ and that there exists $\rho \in \C$ such that $\g(\rho) = \g^*$. One can then define the growth-optimal portfolio $\hX$ using \eqref{eq: num ELM}. Defining $\Op$ via $\Op_t = \g^* t$, and with $\hX(x) \dfn x \hX$, the asymptotics \eqref{eq: asymptotic  ELM} hold.
\end{prop}

\section{\ito \ Markets and Market Time} \label{sec: ito}

As already mentioned in the Introduction, the growth-optimal portfolio is \emph{not} optimal for the problem of minimizing the expected calendar time to reach a wealth level when considering models where the coefficients may change randomly through time. If the objective is somewhat altered into minimizing expected market time, as we shall define below, then the growth-optimal portfolio is indeed optimal. It is our belief that the notion of market time, as it naturally emerges in our paper, has a very clear and natural interpretation and makes deep sense, and is therefore worth studying beyond the context of the questions raised.

To keep the technical details simple, in this section we assume that $S$ is an It\^o process. Later, in Section \ref{sec: ZMPJR}, we shall see how to relax this assumption to more complex models and still keep the main result holding.

\subsection{The set-up}

The dynamics of the discounted asset-prices are:
\begin{equation} \label{eq: ito model}
\ud S^i_t = S^i_t \pare{ a^i_t \ud t + \sum_{j=1}^m \sigma^{i j}_t \ud W^j_t },
\end{equation}
for each $i = 1, \ldots, d$ and $t \in \Real_+$. Here $a = (a^i)_{i =1, \ldots, d}$ is the predictable $d$-dimensional process of excess \textsl{appreciation rates}, $\sigma = (\sigma^{i j})_{i =1, \ldots, d, \, j =1, \ldots, m}$ is a predictable $(d \times m)$-matrix-valued process of \textsl{volatilities} and $W = (W^j)_{j =1, \ldots, m}$ is a standard $m$-dimensional Brownian motion on $\basis$. We let $c \dfn \sigma \sigma^\top$ denote the $(d \times d)$-matrix-valued process of \textsl{local covariances}.

\subsection{Assumptions}

The general Assumptions \ref{ass: general} have a well-described equivalent for the \ito \ market we are considering.

\begin{prop} \label{prop: ass ito}
Assumptions \ref{ass: general} are equivalent to the following:
\begin{enumerate}
  \item There exists a $d$-dimensional predictable process $\rho$ such that, $(\prob \otimes \Leb)$-a.e., $c \rho = a$. (In that case, $\rho = c^{\dagger} a$ where $c^{\dagger}$ is the Moore-Penrose pseudo-inverse of $c$.)
  \item $\int_0^T |\lambda_t|^2 \ud t < \infty$ for all $T \in \Real_+$, where $\lambda \dfn \sigma^\top c^\dagger a$ is the $m$-dimensional \textsl{risk premium} process. (Then, $|\lambda|^2 = \innerw{a}{\cd a} = \innerw{\rho}{c \rho}$.)
  \item $\int_0^\infty |\lambda_t|^2 \ud t = \infty$, $\prob$-a.s.
\end{enumerate}
In this case, it follows that the logarithm of the \num \ portfolio $\hX$ is given by
\begin{equation} \label{eq: num ito}
\log (\hX) = \frac{1}{2} \int_0^\cdot |\lambda_t|^2 \ud t + \int_0^\cdot \lambda_t \ud W_t.
\end{equation}
\end{prop}

It follows from \eqref{eq: num ito} that $\g_t^* \dfn (1 / 2) |\lambda_t|^2$ equals the maximum growth rate at time $t \in \Real_+$ in the given \ito \ market.

As we did in the case of exponential L\'evy markets, we shall use statements (1), (2) and (3) of Proposition \ref{prop: ass ito} in place of the general Assumptions \ref{ass: general} in what follows.

\subsection{Market time}

With the above notation define now, similar to the previous section, the \textsl{market time} process $\Op = (\Op_t)_{t \in \Real_+}$ by setting it equal to the integral over the maximum growth rate, i.e.,
\[
\Op_t \dfn \int_0^t \g^*_s \ud s \ = \ \frac{1}{2} \int_0^t |\lambda_s|^2 \ud s
\]
for $t \in \Real_+$. Observe that, under the validity of statements (1), (2) and (3) of Proposition \ref{prop: ass ito}, we have $\prob[\Op_\infty = \infty] = 1$ as follows from Proposition \ref{prop: ass ito}(3).
As explained in \S \ref{subsec: the problem}, for given $x < \ell$, our aim is to find the wealth process $X \in \X(x)$ that minimizes $\expec \bra{\TOp (X; \ell) }$.

We briefly explain why the problem of minimizing expected market time to reach a wealth level using such a random clock and not calendar time, is natural and worth studying. Consider for simplicity the one-asset case $d = 1$. Then, at any time $t \in \Real_+$, $|\lambda_t|^2 = |a_t / \sigma_t|^2$ is the ``squared signal to noise ratio'' of the asset-price process or more precisely the squared risk premium. When this quantity is small, the opportunities for making profits over those obtainable from the savings account are rather small; on the other hand, when $|\lambda_t|^2$ is large, at time $t \in \Real_+$ an investor has a lot of opportunities to use the favorable fact that the premium for taking risk is high. Stalling to reach the wealth level $\ell$ when opportunities are favorable should be punished more severely, especially for fund managers, and this is exactly what the market time $\Op$ does. From an economic point of view, market time simply conforms with the underlying growth of the market.

\subsection{The main result}

We are ready to present the solution to the optimization problem of \S \ref{subsec: the problem}, both giving an expression for the value function $v$ and showing again that the growth-optimal portfolio is optimal.

\begin{thm} \label{thm: main of ito}
Under the validity of statements (1), (2) and (3) of Proposition \ref{prop: ass ito} for an \ito \ market, and with $\xhat(x) \dfn x \xhat \in \X(x)$,  for $x < \ell$ we have:
\[
v(x; \ell) \ = \ \log \pare{\frac{\ell}{x}} \ = \ \expec \big[ \TOp (\xhat(x); \ell) \big].
\]
\end{thm}
Once again, this last result is a special case of Theorem \ref{thm: gen} that will be presented in the next section.

\section{Market Time in General Semimartingale Markets} \label{sec: ZMPJR}

The purpose of this section is to give a wide-encompassing definition of market time for semimartingale financial markets and to present a general result on the expected market time to reach a given wealth level, of which both Theorem \ref{thm: main of ELM} and Theorem \ref{thm: main of ito} are special cases. We are now in the very general market model described in Section \ref{sec: descr of prob}.

\subsection{Market time} Guided by the discussions and results in both the exponential L\'evy market case of Section \ref{sec: ELM} and the It\^o market case of Section \ref{sec: ito}, it makes sense to define market time as the underlying optimal growth of the market, i.e., the drift part of the logarithm of the growth-optimal portfolio. We shall have to make minimal assumptions for market time to be well-defined; namely, that the drift part of the logarithm of the growth-optimal portfolio \emph{does} exist.

The following result, which is a refined version of Proposition \ref{prop: if 1 then num  is the one}, ensures that the discussions that follow make sense.

\begin{prop} \label{prop: if 1 then num is the one pred}
Under the validity of Assumption \ref{ass: general}(1), further assume that the logarithm of the \num \ portfolio $\hX$ is a special semimartingale and write $\log(\hX) = \Op + M$ for its canonical decomposition, where $\Op$ is a predictable nondecreasing process and $M$ is a local martingale. Then, Assumption \ref{ass: general}(2) is equivalent to:
\begin{enumerate}
  \item[$(2'')$] $\lim_{t \to + \infty} \Op_t = + \infty$, $\prob$-a.s.
\end{enumerate}
\end{prop}

The following slightly strengthened version of Assumptions \ref{ass: general} will enable us to state our general result in Theorem \ref{thm: gen}.

\begin{asss} \label{ass: general plus}
With Assumptions \ref{ass: general} in force, we further postulate that the logarithm of the \num \ portfolio $\hX$ is a special semimartingale.
\end{asss}

Under Assumptions \ref{ass: general plus}, we can write $\log(\hX) = \Op + M$, where $\Op$ is a predictable nondecreasing process and $M$ is a local martingale. We then \emph{define} market time to be the nondecreasing predictable process $\Op$. According to Proposition \ref{prop: if 1 then num  is the one pred}, we have, $\prob$-a.s., $\Op_0 = 0$ and $\Op_\infty = \infty$. This makes $\Op$ a \emph{bona fide} clock.

\subsection{A general result}

In what follows, $\alpha$ will denote a  nonnegative, possibly infinite-valued random variable such that
\begin{equation} \label{eq: alpha defn}
\frac{\Delta \hX}{\hX_-} \leq \alpha.
\end{equation}
Of course, $\alpha$ can be chosen in a minimal way as  $\alpha \dfn \sup_{t \in \Real_+} (\Delta \hX_t / \hX_{t-})$.

\begin{thm} \label{thm: gen}
Let Assumption \ref{ass: general plus} be in force. With the above definition of the market time $\Op$ and a random variable $\alpha$ satisfying \eqref{eq: alpha defn}, we have
\begin{equation} \label{eq: bounds gen}
\log \pare{\frac{\ell}{x}} \ \leq \ v(x; \ell) \ \leq \ \expec \big[ \TOp(\xhat(x); \ell) \big] \ \leq \ \log \pare{\frac{\ell}{x}} + \expec \bra{\log(1 + \alpha)}
\end{equation}
\end{thm}
It is straightforward that Theorem \ref{thm: gen} covers both Theorem  \ref{thm: main of ELM} and Theorem \ref{thm: main of ito} as special cases. For Theorem  \ref{thm: main of ELM}, $\alpha$ is the constant defined in its statement, while for Theorem  \ref{thm: main of ito} we have $\alpha = 0$.

Dividing the inequalities \eqref{eq: bounds  gen} with $\log(\ell)$ throughout, we get the following corollary of Theorem \ref{thm: gen}.

\begin{cor}
In the setting of Theorem \ref{thm: gen}, suppose that $\expec[\log(1 + \alpha)] < \infty$. Then,
\[
\lim_{\ell \to \infty} \frac{v(x; \ell)}{\log (\ell)} \ = \ 1 \ = \ \lim_{\ell \to \infty} \frac{\expec \big[ \TOp(\xhat(x); \ell) \big]}{\log (\ell)}.
\]
\end{cor}
This last result shows that, under some integrability condition on the possible size of the jumps of the logarithm of the growth-optimal portfolio, the problem of possible overshoots vanishes asymptotically when considering increasing wealth levels $\ell$.

\section{Proofs} \label{sec: proof}

Before we embark on proving all the results of the previous sections, we define, in accordance to \eqref{eq: mark upcross defn}, for any \cadlag \ process $X$ and $\ell \in \Real_+$,
\[
\tau (X; \ell) \dfn \inf \set{t \in \Real_+ \such X_t \geq \ell}.
\]
to be the \textsl{first upcrossing calendar time of $X$ at level $\ell$}. It is clear that $\tau (X; \ell)$ is a stopping time and that $\Op_{\tau (X; \ell)} = \TOp (X; \ell)$ for all \cadlag \ processes $X$ and $\ell \in \Real_+$.

\subsection{Proof of Proposition \ref{prop: if 1 then num is the one}}
Recall that the clock $\Op$ satisfies, $\prob [\Op_\infty = \infty] = 1$. Therefore, for any $X \in \X$ and $\ell \in \Real_+$, $\prob[\tau(X; \ell) < \infty] = 1$ is equivalent to $\prob[\TOp(X; \ell) < \infty] = 1$.

Condition $(2')$ of Proposition \ref{prop: if 1 then num is the one} obviously implies Assumption \ref{ass: general}(2). Conversely, assume that Assumptions \ref{ass: general} are in force. For any $n \in \Natural$, pick $X \in \X(1)$ such that, $\prob[\tau^n < \infty] = 1$, where $\tau^n \dfn \tau(X; n)$. Since $X / \hX$ is a nonnegative supermartingale, the optional sampling theorem (see for example \S1.3.C of \cite{MR1121940}) gives:
\[
1 \geq \expec \bra{\frac{X_{\tau^n}}{\hX_{\tau^n}}} \geq n \expec \bra{\frac{1}{\hX_{\tau^n}}}.
\]
It follows that $(1 / \hX_{\tau^n})_{n \in \Natural}$ converges to zero in probability. As $1 / \hX$ is a nonnegative supermartingale, this implies that $\lim_{t \to \infty} (1 / \hX_t) = 0$, $\prob$-a.s., which establishes the result.

\subsection{Proof of Proposition \ref{prop: if 1 then  num is the one pred}}

Under the assumption that the \num \ portfolio $\hX$ is a special semimartingale with canonical decomposition $\hX = \Op + M$, the event equality
\[
\set {\lim_{t \to \infty} \hX_t = + \infty } = \set {\lim_{t \to \infty} \Op_t = + \infty },
\]
which is to be understood in a modulo $\prob$ sense, is a consequence of Proposition 3.21 in \cite{MR2335830}. Then, the result of Proposition \ref{prop: if 1 then  num is the one pred} readily follows in view of Proposition \ref{prop: if 1 then  num is the one}.

\subsection{Proof of Proposition \ref{prop: nec and suf ELM}}

The fact that $\I \cap \Cc = \emptyset$ is equivalent to the existence of $\rho \in \C$ such that $\g(\rho) = \g^* < \infty$, as well as that $\hX$ as defined in \eqref{eq: num ELM} is the \num \ portfolio is a consequence of Lemma 4.1 in \cite{Kar07}, as soon as one also uses the bounded-jump Assumption \ref{ass: bdd levy meas}.

Now, it is straightforward to check that $\g^* = 0$ is equivalent to $\hX$ being a positive local martingale, in which case we have that, $\prob$-a.s., $\lim_{t \to \infty} \hX_t < \infty$. On the other hand, if $\g^* > 0$ then the L\'evy process $\log (\hX)$ is integrable and has strictly positive drift $\g^*$; therefore, $\prob$-a.s., $\lim_{t \to \infty} \hX_t = \infty$. In view of Proposition \ref{prop: if 1 then num is the one}, the result follows.

\subsection{Proof of Proposition \ref{prop: ass ito}}

The fact that (1) and (2) of Proposition \ref{prop: ass ito} are equivalent to the existence of the \num \ portfolio $\hX$, as well as that $\hX$ given by \eqref{eq: num ito}, is a special case of Theorem 3.15 in \cite{MR2335830} --- see also \cite{MR1384360}. Under the validity of (1) and (2) of Proposition \ref{prop: ass ito}, it is straightforward to see that (3) of Proposition \ref{prop: ass ito} is equivalent to $\lim_{t \to \infty} \hX_t = \infty$. Using Proposition \ref{prop: if 1 then num is the one}, the result follows.

\subsection{Proof of Theorem \ref{thm: gen}}
\label{subsec: proof of main ito}

Let $\hL(x) \dfn \log (\hX(x))$. Observe that, since $\Delta \xhat \leq \alpha \xhat_-$,
\begin{equation} \label{eq: bound on log-jumps}
\Delta \Lhat (x) = \log \pare{1 + \frac{\Delta \xhat}{\xhat_-}} \leq \log(1 + \alpha).
\end{equation}
Write $\hL(x) = \log(x) + \Op + M$,
where $M$ is a local martingale.
Let $(\tau^n)_{n \in \Natural}$ be a localizing sequence for $M$.
The estimate \eqref{eq: bound on  log-jumps} gives, for all $n \in \Natural$,
\[
\log (x) + \expec \bra{\Op_{\tau^n \wedge \tau(\xhat(x); \ell )}} = \expec \bra{\hL_{\tau^n \wedge \tau(\xhat(x); \ell )}(x)} \leq \log(\ell) + \expec[\log (1 + \alpha)].
\]
Letting now $n$ tend to infinity and using the monotone convergence theorem, we get
\begin{equation} \label{ineq: ito 1}
\expec \big[ \TOp(\xhat(x); \ell) \big] \leq \log(\ell / x) + \expec[\log (1 + \alpha)].
\end{equation}

\smallskip

Take now any $X \in \X(x)$. If $\prob[\TOp \pare{X, \ell} = \infty] > 0$, we have $\expec \bra{\TOp \pare{X, \ell}} = \infty$ and $\log (\ell / x) \leq \expec \bra{\TOp \pare{X, \ell}}$ is trivial. It remains to consider the case $\prob[\TOp \pare{X, \ell} < \infty] = 1$, or equivalently $\prob[\tau \pare{X, \ell} < \infty] = 1$.

For all $\epsilon \in (0,1)$, define $X^\epsilon := (1-\epsilon) X + \epsilon x$. Then, $X^\epsilon \in \X(x)$ and $\tau \pare{X^\epsilon, \epsilon x + (1 - \epsilon) \ell} = \tau \pare{X, \ell}$. The drift part of the process $L^\epsilon \dfn \log \pare{X^\epsilon}$ is bounded above by $\Op$. Therefore,
\[
L^\epsilon \leq \log(x) + \Op + M^\epsilon
\]
for some local martingale $M^\epsilon$. Let $(\tau^{\epsilon, n})_{n \in \Natural}$ be a localizing sequence for $M^\epsilon$. Since the stopped process $M^\epsilon_{\tau \pare{X, \ell} \wedge \tau^{\epsilon, n} \wedge \cdot}$ is a martingale, we have that
\[
\expec \bra{L^\epsilon_{\tau \pare{X, \ell} \wedge \tau^{\epsilon, n} }} \ \leq \ \log(x) + \expec \bra{\Op_{\tau \pare{X, \ell} \wedge  \tau^{\epsilon, n}} } \ = \ \log(x) + \expec \bra{\TOp \pare{X, \ell} \wedge \Op_{\tau^{\epsilon, n}}}.
\]

Now, $L^\epsilon$ is uniformly bounded from below by $\log(\epsilon x)$. Furthermore, $\uparrow \limn \Op_{\tau^n} = \infty$ holds in a $\prob$-a.s. sense. Therefore, applications of Fatou's Lemma and the monotone convergence theorem will give
\begin{eqnarray*}
  \log(\ell) + \log(1 - \epsilon)\ \leq \ \expec \bra{L^\epsilon_{\tau \pare{X, \ell}}} &\leq& \liminf_{n \to \infty} \expec \bra{L^\epsilon_{\tau \pare{X, \ell} \wedge \tau^{n}}} \\
   &\leq& \log(x) + \liminf_{n \to \infty} \expec \bra{\TOp \pare{X, \ell} \wedge \Op_{\tau^n}} \\
   &=& \log(x) +  \expec \bra{\TOp \pare{X, \ell}}.
\end{eqnarray*}
Sending now $\epsilon$ to zero, we also get $\log (\ell / x) \leq \expec \bra{\TOp \pare{X, \ell}}$ for all $X \in \X(x)$ that satisfy $\prob[\TOp \pare{X, \ell} < \infty] = 1$. This, coupled with \eqref{ineq: ito 1}, finishes the proof.

\subsection{Proof of Proposition \ref{prop: asymptotic  ELM}}

The existence of $\rho \in \C$ such that $\g(\rho) = \g^* < \infty$ follows from Lemma 4.1 in \cite{Kar07} in view of $\I \cap \Cc \neq \emptyset$. Note that the finiteness of $\g^*$ is straightforward from the defining equation \eqref{eq: growth rate} for $\g$.

Call $\hL \dfn \log(\hX)$. For each $n \in \Natural$, let
\[
\hL^n \dfn \hL - \sum_{t \leq \cdot} (\Delta \hL_t) \indic_{\{ \Delta \hL_t > n \}}.
\]
Then, $\hL^n$ is a L\'evy process and we can write
\[
\hL^n_t = \g^n t + M^n_t
\]
for all $t \in \Real_+$, where $M^n$ is a L\'evy martingale and $\uparrow \limn \g^n = \g^* > 0$. Then,
\[
\expec[\TOp (\hX(x); \ell)] = \g^* \expec[\tau (\hX(x); \ell)] \leq \g^* \expec \bra{\tau \pare{\hL^n(x); \log(\ell)}} \leq \frac{\g^*}{\g^n} \pare{\log \pare{ \frac{\ell}{x}} + \log(1 + n)},
\]
holds for all $n \in \Natural$ such that $\g^n > 0$, where the last inequality follows along the same lines of the proof of \eqref{ineq: ito 1}. It then follows that
\[
\limsup_{\ell \to \infty} \frac{\expec[\TOp (\hX(x); \ell)]}{\log (\ell)} \leq \frac{\g^*}{\g^n}
\]
holds for all $n \in \Natural$ such that $\g^n > 0$. Since $\uparrow \limn \g^n = \g^* > 0$, sending $n$ to infinity in the last inequality we get
\[
\limsup_{\ell \to \infty} \frac{\expec[\TOp (\hX(x); \ell)]}{\log (\ell)} \leq 1.
\]
Of course, in view of the bounds \eqref{eq: bounds  gen} of Theorem \ref{thm: gen}, we always have
\[
1 = \lim_{\ell \to \infty} \frac{v(x; \ell)}{\log (\ell)} \leq \liminf_{\ell \to \infty} \frac{\expec[\TOp (\hX(x); \ell)]}{\log (\ell)},
\]
which completes the proof.

\bibliographystyle{siam}
\bibliography{expec_time_min}

\begin{thebibliography}{10}

\bibitem{MR929084}
{\sc P.~H. Algoet and T.~M. Cover}, {\em Asymptotic optimality and asymptotic
  equipartition properties of log-optimum investment}, Ann. Probab., 16 (1988),
  pp.~876--898.

\bibitem{MR520737}
{\sc D.~C. Aucamp}, {\em An investment strategy with overshoot rebates which
  minimizes the time to attain a specified goal}, Management Sci., 23
  (1976/77), pp.~1234--1241.

\bibitem{MR1849424}
{\sc D.~Becherer}, {\em The numeraire portfolio for unbounded semimartingales},
  Finance Stoch., 5 (2001), pp.~327--341.

\bibitem{Black76}
{\sc F.~Black}, {\em Studies of stock price volatility changes}, in Proceedings
  of the 1976 Meetings of the Business and Economics Statistics Section,
  American Statistical Association, Berkeley, Calif., 1976, pp.~177--181.

\bibitem{MR0135630}
{\sc L.~Breiman}, {\em Optimal gambling systems for favorable games}, in Proc.
  4th Berkeley Sympos. Math. Statist. and Prob., Vol. I, Univ. California
  Press, Berkeley, Calif., 1961, pp.~65--78.

\bibitem{MR1724568}
{\sc S.~Browne}, {\em Reaching goals by a deadline: digital options and
  continuous-time active portfolio management}, Adv. in Appl. Probab., 31
  (1999), pp.~551--577.

\bibitem{MR2284490}
{\sc M.~M. Christensen and K.~Larsen}, {\em No arbitrage and the growth optimal
  portfolio}, Stoch. Anal. Appl., 25 (2007), pp.~255--280.

\bibitem{MR1384360}
{\sc F.~Delbaen and W.~Schachermayer}, {\em The existence of absolutely
  continuous local martingale measures}, Ann. Appl. Probab., 5 (1995),
  pp.~926--945.

\bibitem{MR1842286}
{\sc H.~F{\"o}llmer and P.~Leukert}, {\em Quantile hedging}, Finance Stoch., 3
  (1999), pp.~251--273.

\bibitem{MR872458}
{\sc D.~Heath, S.~Orey, V.~Pestien, and W.~Sudderth}, {\em Minimizing or
  maximizing the expected time to reach zero}, SIAM J. Control Optim., 25
  (1987), pp.~195--205.

\bibitem{HeSu}
{\sc D.~Heath and W.~Sudderth}, {\em Continuous-time portfolio management:
  Minimizing the expected time to reach a goal}.
\newblock Unpublished manuscript, 1984.

\bibitem{MR2335830}
{\sc I.~Karatzas and C.~Kardaras}, {\em The num\'eraire portfolio in
  semimartingale financial models}, Finance Stoch., 11 (2007), pp.~447--493.

\bibitem{MR1121940}
{\sc I.~Karatzas and S.~E. Shreve}, {\em Brownian motion and stochastic
  calculus}, vol.~113 of Graduate Texts in Mathematics, Springer-Verlag, New
  York, second~ed., 1991.

\bibitem{Kar07}
{\sc C.~Kardaras}, {\em No-{F}ree-{L}unch equivalences for exponential {L}\'evy
  models}.
\newblock To appear in Mathematical Finance, 2007.

\bibitem{KarPla07}
{\sc C.~Kardaras and E.~Platen}, {\em On the semimartingale property of
  discounted asset-price processes in financial modeling}.
\newblock submitted for publication, 2008.

\bibitem{MR0090494}
{\sc J.~L. Kelly, Jr.}, {\em A new interpretation of information rate}, Bell.
  System Tech. J., 35 (1956), pp.~917--926.

\bibitem{Long90}
{\sc J.~B.~J. Long}, {\em The num\'eraire portfolio}, Journal of Financial
  Economics, 26 (1990), pp.~29--69.

\bibitem{MR2194899}
{\sc E.~Platen}, {\em A benchmark approach to finance}, Math. Finance, 16
  (2006), pp.~131--151.

\bibitem{MR1739520}
{\sc K.-I. Sato}, {\em L\'evy processes and infinitely divisible
  distributions}, vol.~68 of Cambridge Studies in Advanced Mathematics,
  Cambridge University Press, Cambridge, 1999.

\end{thebibliography}
\end{document}